\documentclass[12pt,preprint]{aastex}

\slugcomment{Preprint, \today}

\shorttitle{High-Arctic Mountain Sites}
\shortauthors{Steinbring et al.}

\input epsf
\def\plotone#1{\centering \leavevmode
\epsfxsize=0.6\columnwidth \epsfbox{#1}}
\def\plotonenarrow#1{\centering \leavevmode
\epsfxsize=0.5\columnwidth \epsfbox{#1}}
\def\plotonewide#1{\centering \leavevmode
\epsfxsize=0.7\columnwidth \epsfbox{#1}}

\DeclareMathAlphabet{\mathscr}{OT1}{pzc}{m}{it}

\begin{document}

\title{First Assessment of Mountains on Northwestern Ellesmere Island, 
Nunavut, as Potential Astronomical Observing Sites}

\author{Eric Steinbring\altaffilmark{1}, Ray Carlberg\altaffilmark{2}, Bryce 
Croll\altaffilmark{2} Greg Fahlman\altaffilmark{1}, Paul 
Hickson\altaffilmark{3}, Liviu Ivanescu\altaffilmark{4}, Brian 
Leckie\altaffilmark{1}, Thomas Pfrommer\altaffilmark{3} \& Matthias 
Schoeck\altaffilmark{1}}

\altaffiltext{1}{Herzberg Institute of Astrophysics, National Research
Council Canada, Victoria, BC V9E 2E7, Canada}

\altaffiltext{2}{University of Toronto, Dept. of Physics and Astronomy, Toronto, ON M5S 3H4, Canada}

\altaffiltext{3}{University of British Columbia, Dept. of Physics and Astronomy, BC V6T 1Z1, Canada}

\altaffiltext{4}{University of Montreal, Dept. of Physics and Astronomy, QC H3C 3J7, Canada}

\begin{abstract}

Ellesmere Island, at the most northerly tip of Canada, possesses the highest 
mountain peaks within 10 degrees of the pole. The highest is 2616 m, with many 
summits over 1000 m, high enough to place them above a stable low-elevation 
thermal inversion that persists through winter darkness. Our group has 
studied four mountains along the northwestern coast which have the additional 
benefit of smooth onshore airflow from the ice-locked Arctic Ocean. We deployed 
small robotic site testing stations at three sites, the highest of which is over 
1600 m and within 8 degrees of the pole. Basic weather and sky clarity data for
over three years beginning in 2006 are presented here, and compared with 
available nearby sea-level data and one manned mid-elevation site. Our results 
point to coastal mountain sites experiencing good weather: low median wind 
speed, high clear-sky fraction and the expectation of excellent seeing. Some 
practical aspects of access to these remote locations and operation and 
maintenance of equipment there are also discussed.
\end{abstract}

\keywords{site testing}

\section{Introduction}\label{introduction}

The cold, dry, dark winter skies of the earth's polar regions are well suited 
for astronomy. Smooth airflow is aided by a highly stratified atmosphere with 
strong, stable low-elevation thermal inversions that persist during the long 
night. And in particular, extreme latitudes enjoy clear skies and mild winds 
associated with regions of polar high pressure. In Antarctica, these conditions 
have encouraged significant focus on the central high-elevation ice plateau 
\citep[e.g.][and references therein]{Saunders2009, Lawrence2010}. This has been aided by a year-round manned 
presence at the South Pole (latitude 90 South, 2835 m elevation a.s.l.), and 
more recently Dome C (76S, 3260 m). A permanent station at Dome A (80S, 4200 m), 
the highest elevation on the plateau, is planned \citep{Gong2010}. In the
arctic, as an analogue to the antarctic plateau, the summit of the Greenland 
icecap (72N, 3200m) has gained interest \citep{Andersen2006}.

An exciting development in polar astronomy was experimental proof of excellent 
free-atmospheric seeing at Dome C, Antarctica, with blurring induced by
the high atmosphere of only 0\farcs27 at $V$ \citep{Lawrence2004}. The important
implication is that a telescope sited in such conditions can be quite small yet still
very productive; a senstivity advantage allowing even a modest, 2-m class 
optical/near-infrared observatory to be competitive with larger facilities
at the current best mid-latitude sites of Hawaii and Chile \citep{Lawrence2009}.
However, a concern is that strong boundary-layer turbulence is found to blanket the ice plateau. At Dome C the median seeing 8.5 m above the ice surface is 1\farcs8$\pm$0\farcs8 \citep{Agabi2006}, and to take advantage of the free atmospheric seeing there would require a telescope be mounted on a 30-m high tower \citep{Saunders2008}.

An alternate, and more traditional site would be a polar mountain, which would not be subject to the same boundary-layer issues as a flat glacial plateau. Ranges in Canada's Eastern Arctic archipelago are of particular note, having many 
summits between 1000 m and 2000 m that project into the free atmosphere. None are 
as high as the antarctic or Greenland icecaps, but Mount Barbeau (82N, 2616 m) on Ellesmere Island, Nunavut, is higher than peaks in northern Greenland, and is almost as high as Cerro Pachon 
(2715 m) in Chile. And in cold, dry air the effective altitudes of the arctic peaks is 
increased. Being located as far north as a latitude of 82 degrees they offer the 
advantage of a longer night, superior to sites further from the poles. This improves the prospects for long time-series observations 
of, for example, planetary transit searches \citep{Rauer2008}. The other distinct 
advantage of a mountain site over an icecap is a solid, rock foundation on which 
to install a telescope.

Continuous meteorological records going back over 50 years are available for the
two year-round manned stations on Ellesmere Island: Alert (82N) and Eureka
(80N). All of Ellesmere Island is a polar desert, the coldest and driest in the High Arctic, with annual precipitation less
than 9 cm, occuring primarily in summer. Daily climate normals are -40C for winter and
5C in summer. Weather follows a remarkably consistent pattern associated with
the arctic polar vortex, and a strong surface-based thermal inversion persists
in the dark: September to March. So during this time the free atmosphere is
prevented from mixing, both with lesser latitudes (outside the vortex) and lower
altitudes (beneath the surface-based thermal inversion). The arctic polar vortex
is not quite as uniform as the antarctic one, though, being generally more elongated
and not perfectly centered on the pole, but the region it encircles contains
comparably cold, dry, and stable air.

Both Alert and Eureka are accessible year-round by air, and Eureka is supplied
annually by ship in summer. Alert is a Canadian military outpost,
while Eureka is primarily operated by Environment Canada (EC), a civilian government
department. Commercial chartered flights are available to Eureka, generally as
combination passenger/cargo operations using Boeing 727 or 737 or smaller
turboprop aircraft, all modified for gravel runways. Bushplanes and helicopters
supporting scientific research on Ellesmere Island are marshalled at Eureka by
Natural Resources Canada through the Polar Continental Shelf Program (PCSP).

Eureka is the furthest north that geosynchronous communications satellites can
reliably be used. In fact, Alert is linked to the south through Eureka via six
microwave repeater towers. Each is on a high point in a chain along the mountains
running the length of the island. The highest of these repeaters (called ``Grant", 1256 m) is nearby Alert, with the others along the southeastern flank of the
mountains, all accessed by helicopter only. The last is on Black Top Ridge (825
m), to the east of Eureka. The satellite tranceiver is on a 610-m ridge
accessible by road, to the west of Eureka. At
the top is the Polar Environment Atmospheric Research Laboratory (PEARL),
operated by the Canadian Network for the Detection of Atmospheric Change
(CANDAC). In 2005/06 this university collaboration redeveloped the previous EC
facility, which was decomissioned in the 1990s. It has reliable power - supplied by diesel
generators in Eureka - and broadband satellite communications.

This paper outlines our study of selected mountains on the northwestern coast of
Ellesmere Island for their use as sites for telescopes, primarily in the optical and near infrared. In 2006 through 2009
four candidate mountain sites were selected for study, and on three of those,
meteorological and sky information were gathered \citep{Steinbring2008,
Wallace2008}. The program is described in Section~\ref{testing}, including deployment of site-testing instrumentation in the arctic. The collected data are presented in Section~\ref{weather} and analyzed in comparison with data from nearby sea-level stations, including Alert and Eureka, one mid-elevation manned
site at PEARL, and the highest possible elevation peak on Ellesmere Island. A summary of the results follows in Section~\ref{discussion}, with discussion focussing on the practicalities of the sites and the possibility of future development.

\section{Site Testing on Ellesmere Island Mountains}\label{testing}

A good observing site requires clear skies and a dry, calm atmosphere with the
smoothest possible airflow. This generally leads to the selection of a mountain near
the sea coast from which the prevailing winds blow with a desert near its base.
A high, isolated peak will reduce the influence of upwind terrain. Plus, a higher elevation will result in less air
column, lowering precipitable water vapour and improving the expectation of good
seeing \citep{Racine2005}. For arctic astronomy, being close to the pole results in lower sun
elevations, providing a longer night. Eureka experiences 4020 hours per year when the sun is below the horizon; 2419 hours of nautical twilight or darker (sun-angle $<-12$ degrees) and 1461 hours of astronomical darkness ($<-18$ degrees). It is also at sufficiently high latitude to place it within the northern auroral hole; when seen from Eureka, aurora appear low in the
southern sky.  And a preliminary satellite analysis of Ellesmere Island
in winter also pointed to higher clear-sky fraction with increasing latitude and elevation. This will be discussed further in Section~\ref{weather}.

Thus, purely based on geography, the most desirable arctic mountain site will maximize
the following four conditions:

\begin{itemize}
\item{Greatest latitude}
\item{Highest elevation}
\item{Nearest to coast}
\item{Most isolated peak}
\end{itemize}

These criteria quickly point to the mountains of northern Ellesmere Island.
They are unique relative to other arctic sites in being true mountains with high
prominances, as well as being very near the coast. By comparison, nunataks
(rocky areas jutting up from an ice field) although more common on Ellesmere Island,
and in some cases significantly higher, would provide less relief above a
relatively flat ice surface. These are also more prevalent further inland, not
near the coast.

\subsection{Preliminary Downselection}

As a first step in selecting sites for study, the fourteen highest peaks on
Ellesmere Island with latitudes of 80N or greater, within 100 km of the coast
and not within an ecologically protected area were identified from digital
elevation maps (DEMs).

Next, high-resolution satellite imagery was studied. Mountains nearer the coast had clearly visible exposed rock along their flanks in summertime. So these mountains were 
considered more
accessible but not necessarily having the best sky conditions. For none of the mountains was it possible to
determine from either DEMs or images whether a suitably flat area was
available for landing a helicopter.

Aircraft are stationed at Alert and Eureka. From there, a one-way flight of 300
km is effectively a limiting range for most helicopters, requiring refueling for
the return journey. So near sea level, an area for landing and takeoff of
fixed-wing aircraft is also needed - an ``unimproved" airstrip sufficient for a bushplane - 
allowing the cacheing of fuel for helicopter
trips and support for a temporary base camp. The availability of suitable
locations with unimproved airstrips was discussed with arctic logistics experts
familiar with Ellesmere Island. Essentially all of Ellesmere Island could be
accessed from Eureka and Alert, although a requirement of access from one or the other,
not both, was seen as more efficient. This also pushed towards selecting a group
of mountains that could be accessed from the same base camp.

From the initial sample of 14 mountains, a downselection was made based on the
following criteria:

\begin{itemize}
\item{Availability of rock near summit, especially desirable if exposed during summer months}
\item{Possibility of a nearby camp, with a suitable unimproved airstrip}
\item{Proximity to a major research base}
\item{Ability to get permits for testing and potentially for construction}
\end{itemize}

\begin{figure}
\plotonewide{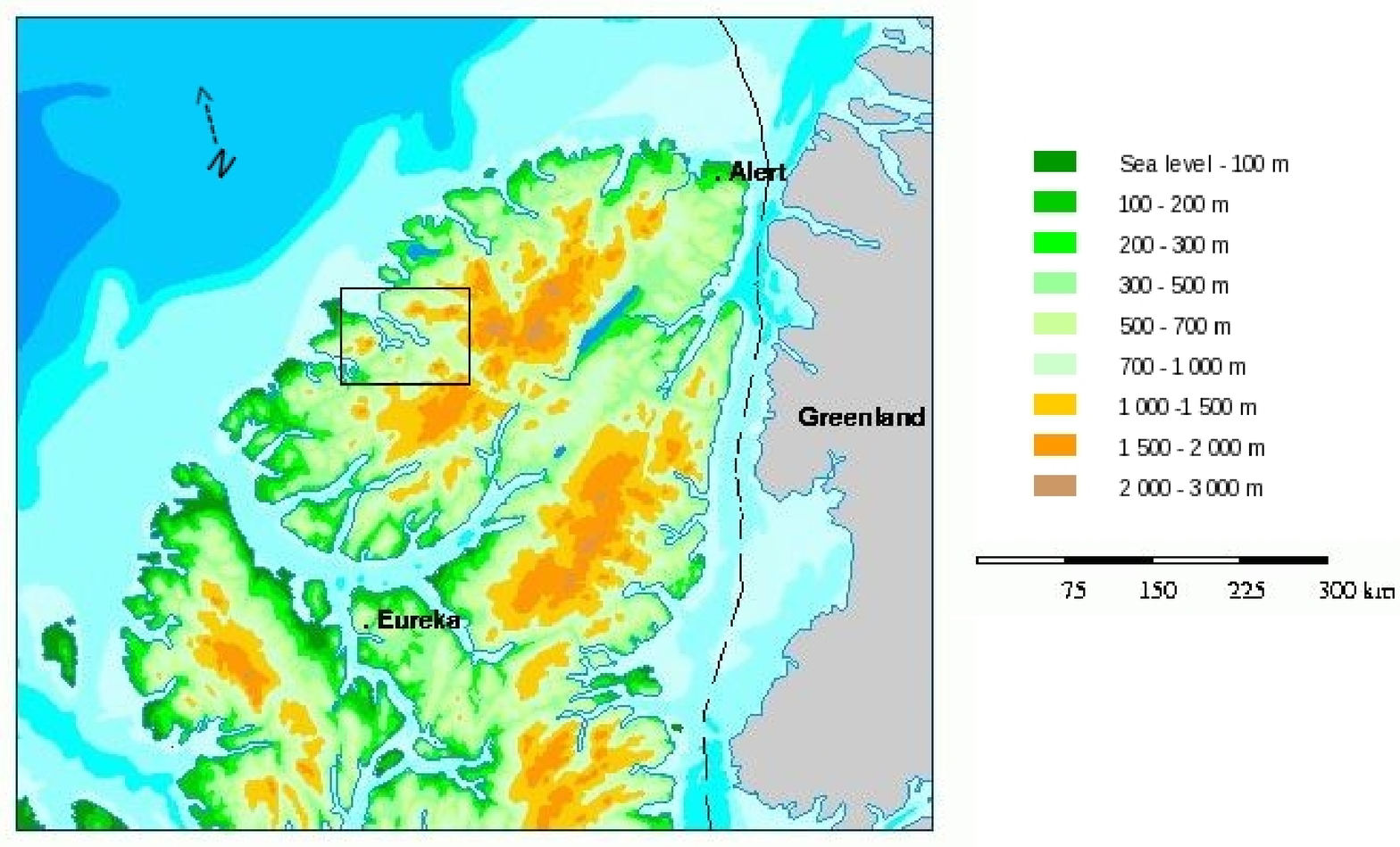}\\
\plotonewide{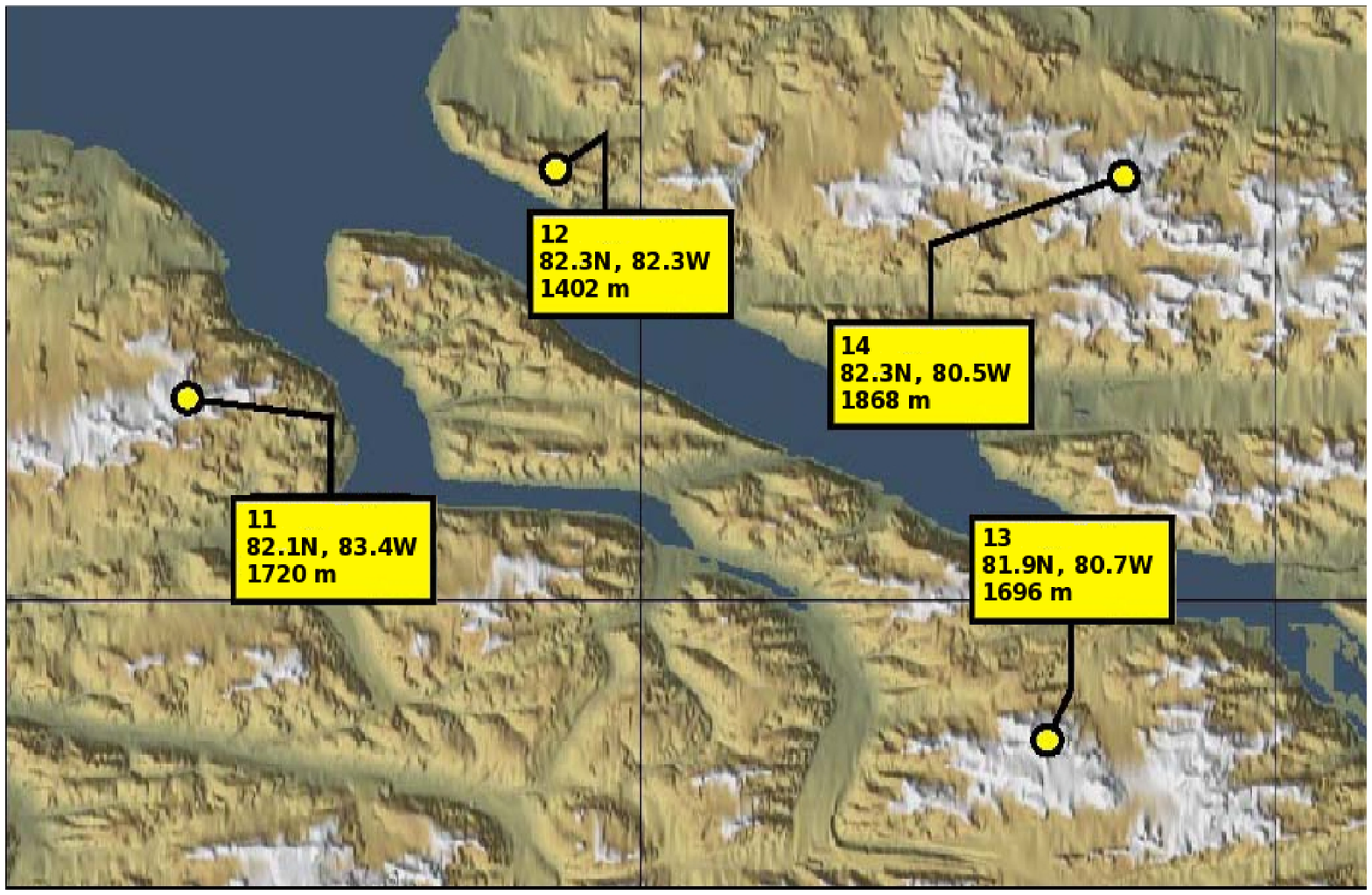}
\caption{Topographic map of Ellesmere Island. The study region is outlined, and below is a higher-resolution map of the selected mountain sites; four isolated peaks over
1000 m within 100 km of the coast. Here white refers to elevations over 1500 m.}
\label{figure_topographic_map}
\end{figure}

The final consideration has already been partly addressed by avoiding the
ecologically protected Quttinirpaaq national park. Further discussion of the
process for obtaining permits for scientific research in the Territory of
Nunavut will follow in Section~\ref{discussion}. The outcome was to focus on the four most northerly mountains in the sample (see Figure~\ref{figure_topographic_map}).
 Two of these are close to the coast,
with two further inland. The inland peaks are somewhat higher (moutain 14 reaches almost 1900 m). All could be accessed by helicopter from a single base camp,
supported from Eureka, roughly 250 km to the south.

\subsection{Aircraft Access to Remote Locations}

Although access to these mountainous locations is possible from spring through
fall, the most efficient period is during mid-summer, when daylight lasts 24
hours and nearby aircraft traffic due to other scientific programs is at its
maximum. At this time there has already been significant melting of snow cover,
allowing well-cleared landing sites.

The mountain sites themselves must be accessible by helicopter, and to minimize travel the unimproved airstrip and camp should be nearby. The northern
fiords of Ellesmere Island present a challenge in this regard, because they can
be subject to fog due to onshore winds over patches of open water. This has
nothing to do with the quality of the sites themselves, since the fog is
confined to the ground, but it restricts travel between the sites and camp and
can persist for several days. It is also, therefore, a safety concern as it
could lead to a situation where personnel become stranded on a mountain peak. We investigated three possible camp locations nearby the study sites, and
used two of those. The best was the most inland, another was along the coast and closer to the sites, and a third - not used - is very near mountain 11.
Since we knew in advance which camp location would be used, some fuel was cached
there ahead of time, reducing weight on the bushplane when making camp.

To reach remote research sites in the Canadian arctic a de Havilland
Twin Otter bushplane is typically used. It is a twin-engined turboprop
with large balloon tires that can land on almost any rough airstrip of 200 metres
or more. It has a range of 1300 km and takeoff cargo weight of 1100 kg. A
larger cousin, the Buffalo, has a payload of over 8000 kg. A commonly used
helicopter in the arctic is a Bell 206, a medium-sized single-turbine machine.
It has a maximum capacity of either 5 people (including pilot) or about 300 kg near sea-level. A variant with either 7 seating positions or more room for
cargo - the 206L ``Longranger" - is available, with a similar capacity. Both variants
have a range of about 500 km. A somewhat larger machine, the Bell 407, can
carry roughly double the load within a similar range.

For helicopters, ceiling and and range decrease sharply with increasing loads.
PCSP contracts both Twin Otters and Longranger helicopters for research
expeditions in the arctic. It does not normally contract helicopters in the
medium to heavy-lift category (lifting 1000 kg to over 5000 kg), but they do
operate in the the arctic. Notably, in 2006 a
Sikorsky heavy lift helicopter was deployed to recover the BLAST balloon-borne
telescope after touchdown on Victoria Island.

\subsection{Helicopter-Deployable Instrumentation}

The size of helicopter used quickly defines what instrumentation is possible at
the sites. The autonomous site-testing stations we deployed, which we have termed ``Inuksuit" (plural of inuksuk, the Inuktitut word for the iconic stone waymarkers of the North) are discussed in \cite{Steinbring2008}. These battery-powered stations measured
basic weather conditions: air temperature, barometric pressure, relative
humidity, wind speed and direction. They also employed a wide-field camera for
taking images of the horizon in a fixed direction.  Images were taken hourly
when batteries were sufficiently charged, typically lasting a few days; a full
charge would last about two weeks. Weather records consist of hourly averages from
one-minute samples taken at all times. Once per day, when the batteries were
sufficiently charged, the stations relayed weather and health-status information
via the Iridium satellite network. Because batteries were charged with a wind
turbine - a solar array would be useless in the dark, and a sufficiently large
bank of batteries to last through the winter impractical - an inherent bias in
this approach is to preferentially obtain sky images during windy periods. This
bias can be characterized though, as wind speed is recorded, and the system
performance well understood. To improve on the state of battery charge, we
experimented with a small commercial methanol-based fuel cell.

As a first step in investigating the optical seeing quality of arctic
mountain sites, a lunar scintillometer, the Arctic Turbulence Profiler (ATP) has
been developed \citep{Hickson2010}. At the moment
it is being tested at PEARL, deployed alongside two Sonic Detection and Ranging
systems (SODARs). The ATP is designed to be deployed by Bell 206L helicopter later at
one of the remote sites. Both it and the SODARs measure turbulence profiles in
the planetary boundary layer, within the lower kilometre of the atmosphere. They
are not sensitive to high-level turbulence. Another seeing-monitor device is a
Differential Image Motion Monitor (DIMM) which measures the overall image
quality, but provides no information on what regions of the atmosphere dominate
the degradation. It can be combined with a multi-aperture seeing sensor (MASS),
which provides information on the height distribution of the image degradation,
but is insensitive to ground layer effects. Both MASS and DIMM measurements can
be taken with the same small ($\sim 0.3$ m diameter) telescope, but this is
already quite complicated for deployment at a remote site accessible only by helicopter. We have begun investigating operation of a very small (0.1 m) telescope at our sites, which which will be the subject of a later paper.

\subsection{Selected Mountain Sites}

We opted to employ the helicopter and not fix-winged aircraft for detailed
scouting of the sites. The hourly cost is approximately the same, and there is
a significant advantage to being able to put the helicopter down immediately on
a suspected good site and quickly verify that a landing is possible.

The helicopter pilot has complete control over selecting landing sites, and
safety of all aboard is the forefront concern. There were no satellite images of
the sites at sufficient resolution to pre-determine landing sites, whether on
rock or snow. In general we found that the best approach was to thoroughly
discuss ahead of time with the pilot what locations we considered desirable.  It
is clear that they kept in mind what characteristics are needed while we
performed an aerial search.

In the 2006 field season we installed Inuksuit stations on the two mountains
closer to the coast (one station each on 11 and 12). We selected flat rocky areas as high as
possible and near the coast. They are along the mountain flanks, below the peaks themselves. One is at 1098 m near peak 11 (designated Site 11A). The other is near 
peak 12, at
777 m (12A). Both are on the leading edge of long ridges running down to the
northwest from the main peaks. That they are rocky and relatively snow-free made
helicopter access easier and the solid footing of the stations more assured. The
true summits of both of these mountains were scouted and it was determined that
they did not provide safe landing sites. They were too sharply peaked to land
and so heavily snow-covered that it was not clear how the station could be
anchored. Both of the higher peaks (13 and 14) were also thoroughly scouted.
Mountain 13 is further inland, the furthest from the initial two sites. Mountain 14
looked the more promising of the two, having a broad saddle and a long
snow-covered ridge intersecting the summit, as well as a flatter peak.

In 2007, data from the first year were retrieved, maintenance was carried out on
stations at Sites 11A and 12A, a third site within the snow-covered saddle of mountain 14 was selected (Site 14A), and a station installed there. Further maintenance trips occured in 2008, when a small
fuel-cell electrical generator was installed at Site 11A, and again in 2009. Measurements
of snow depth at Site 14A were made employing a conventional method using an
avalanche probe. We also dug bore holes using a power ice-auger. The snow there is about 1 to 3 m deep. The ice below is hard in places, but in layers,
which can be bored through with the ice auger. Further tests could be pursued by
digging deeper bore holes with a longer ice auger. Based on the surrounding
terrain, rock could be just a few metres below the ice, depending on the
location.

\subsection{Coastal Mountain and Sea-Level Comparisons}

Besides Eureka and Alert, the only other meteorological station on the
northwestern coast of Ellesmere Island at the time of our downselection in 2006 was at
Ward Hunt Island, at sea level. Later sea-level data were taken by other groups
during the following two years. One station was at Cape Alfred Ernest, starting
in 2007, which was later moved to Milne Glacier in 2008. Both locations are
within 50 km of our stations along the coast. The meteorological instrumentation
at Cape Alfred Ernest, Milne Glacier, and Ward Hunt Island are all essentially
identical to ours; obtained at 2 m height from ground, whereas ours are closer
to 1.5 m. Also, in 2006 the PEARL facility was being being re-commissioned. A standard meteorological tower (at 6 m above ground) was added.

The available data from nearby coastal sites provide useful overlap. Comparisons
can also be made with long-term monitoring at Eureka and Alert. A recently
published climatology for Eureka goes back over 50 years \citep{Lesins2009}.
Apart from sea level stations, PEARL provides records at 616 m elevation, which
provides a comparison to our stations at 777 m, 1098 m, and 1639 m, bracketing the peak of the atmospheric thermal inversion.
Table~\ref{table_sites} gives the locations of the sites and comparisons used in
our study.

\section{Observed Weather and Sky Conditions}\label{weather}

In general, site-testing requires a multi-year dataset, because year to year
fluctuations in nighttime conditions for a given month can be large, and may
give a misleading impression. For the arctic, where night lasts several months per year, we consider that data over a significant fraction of one winter period should be minimally sufficient to confirm the expected weather
patterns in darkness. But collecting even this comparatively small amount of data has not been an easy task. In the harsh conditions of the arctic, with autonomous stations
serviced only in summer, significant gaps in the datasets - weeks to months - are bound to occur. Batteries can get too cold to power the electronics. Thermometer and
hygrometer housings are particularly prone to icing and damage. Even with good
electrical grounding, buildup of static electricity in dry air is a problem for
electronics, sometimes with dramatic consequences. The Inuksuit station at Site
12A was destroyed by this in 2007, burning out much of the electronics, including the weather
data-logger. Most of the PEARL data prior to winter 2008 is unusable due to
icing, and related mechanical problems. In separate storms during 2008, the entire station at Cape Alfred Ernest
was blown down, as was our Inuksuit station at Site 14A.

\begin{figure}
\plotonewide{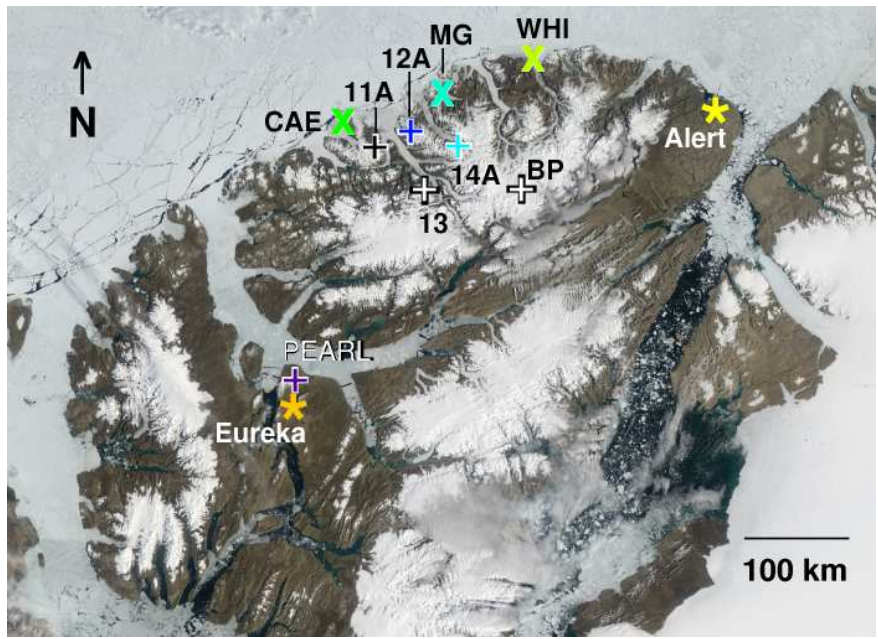}
\caption{Map of Ellesmere Island study locations.  Also indicated
are coastal sea-level stations at Cape Alfred Ernest (CAE), Milne Glacier (MG)
and Ward Hunt Island (WHI); and the highest elevation on the island, Barbeau Peak (BP). Symbols used here are
maintained in plots to follow.}
\label{figure_ellesmere_map}
\end{figure}

\begin{figure}
\plotonewide{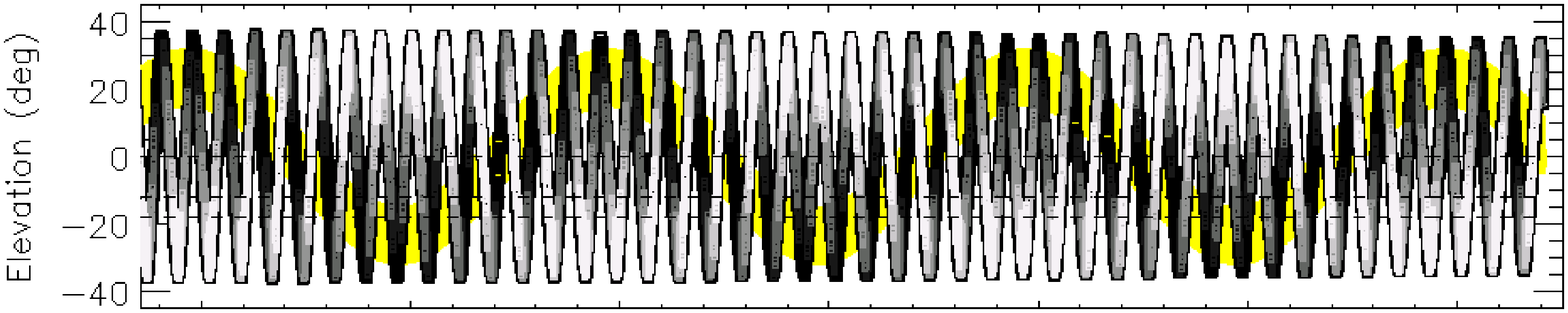}\\
\plotonewide{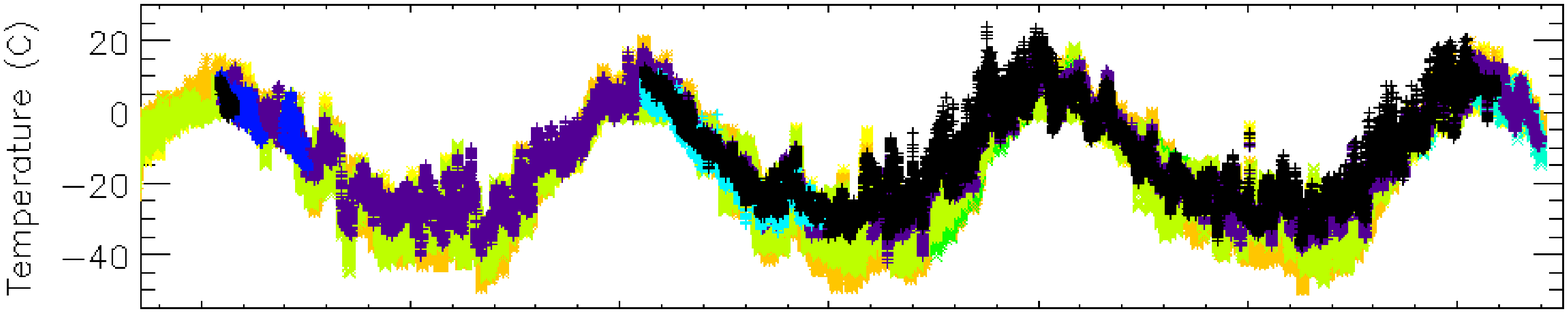}\\
\plotonewide{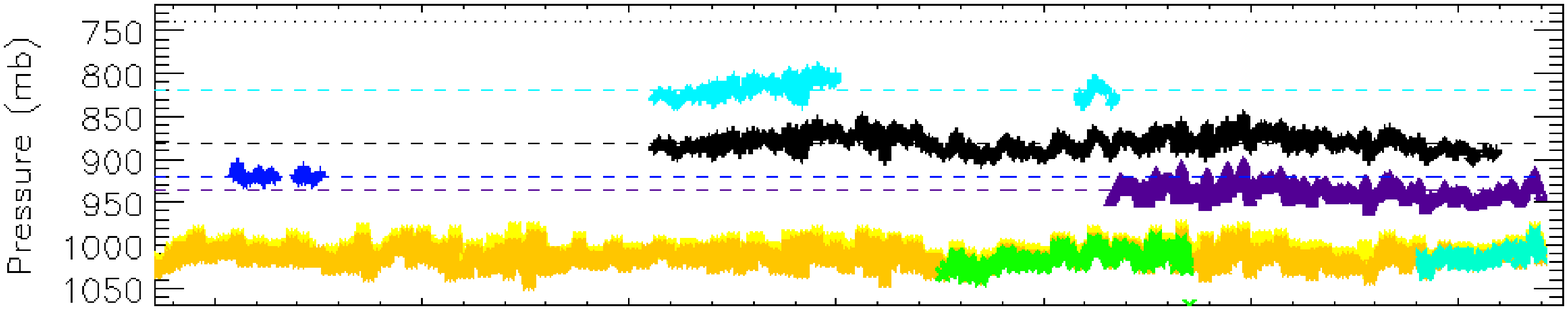}\\
\plotonewide{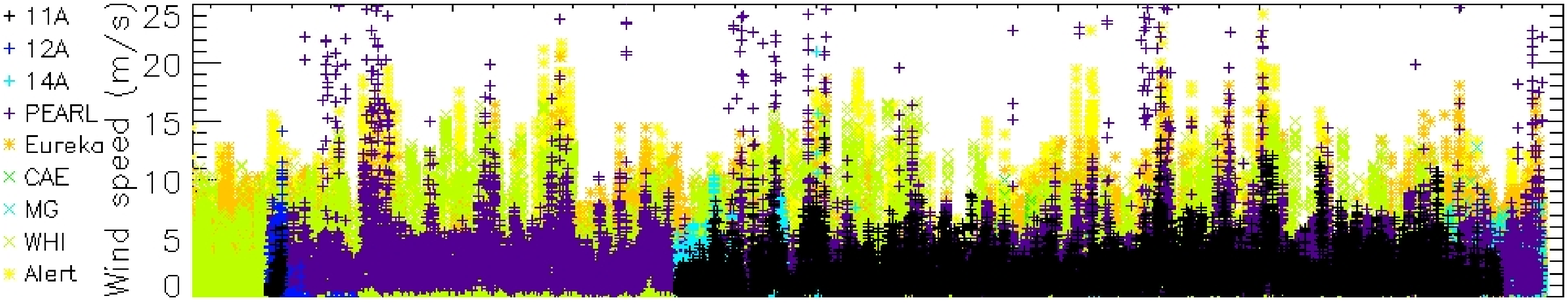}\\
\plotonewide{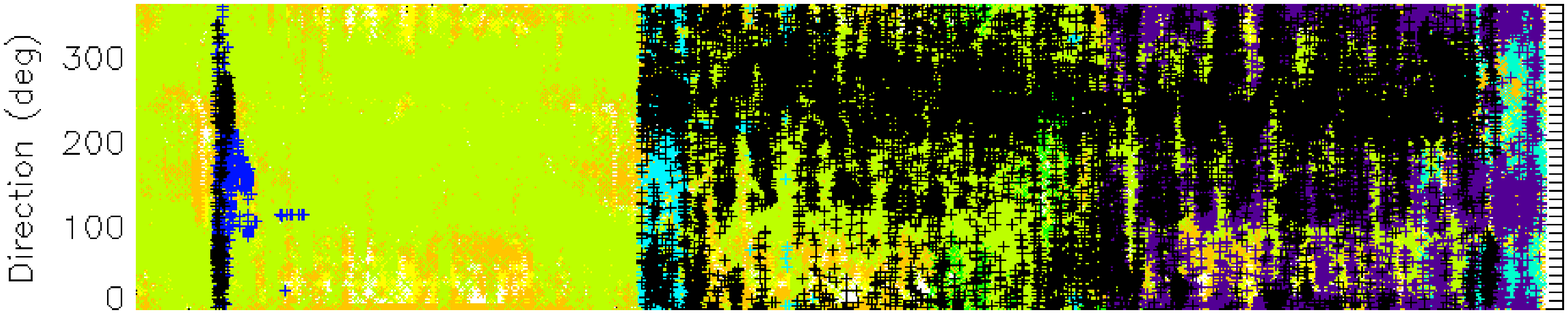}\\
\plotonewide{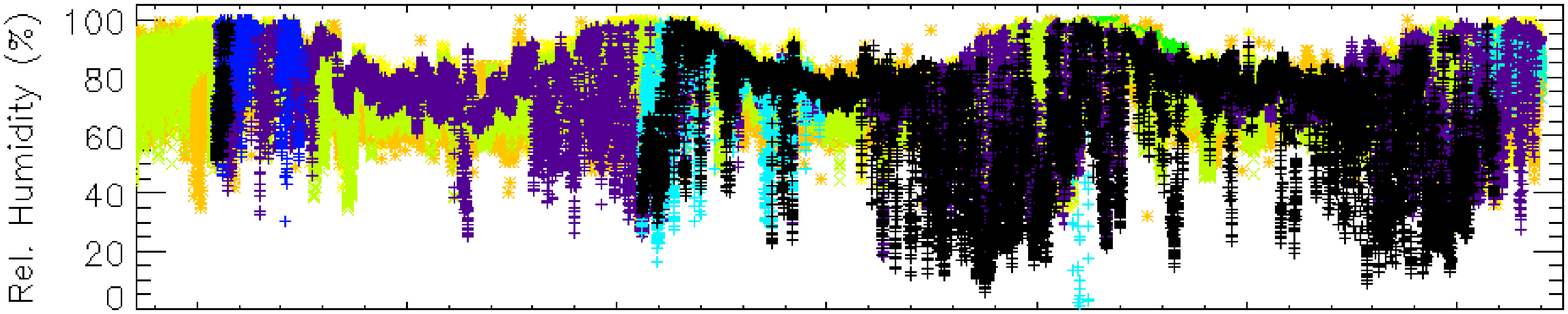}\\
\plotonewide{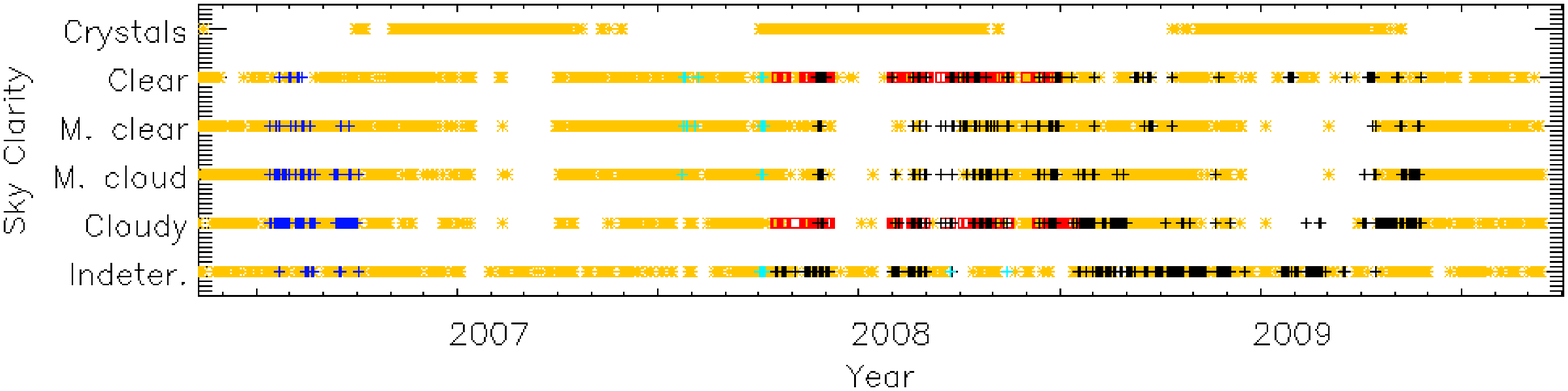}
\caption{Plots of meteorological and sky-clarity data for all stations, starting in July 2006
and continuing until October 2009. Symbols are the same as in Figure~\ref{figure_ellesmere_map}, red squares indicate satellite sky-clarity assessments discussed in Section~\ref{clearsky}. A
plot of sun elevation above the horizon for 80 deg North latitude is shown in
the top panel; moon phase and elevation is also indicated; white is 90\% or
greater illumination, shaded increasingly darker grey down to 10\%, beyond which is
black.}
\label{figure_weather}
\end{figure}

Figure~\ref{figure_weather} shows plots of all meteorological and sky quality
data for the study locations for over three years beginning in July 2006; symbols
follow those used in Figure~\ref{figure_ellesmere_map}. Corrupted and suspect
records, for example, long periods of exactly zero wind speed or wildly varying
barometric pressure have been deleted. The Ward Hunt Island station does not have
a barometer. Each measured parameter will be discussed in the following
sections.

\subsection{Nighttime Thermal Inversion}\label{inversion}

The second and third from top panels of Figure~\ref{figure_weather} are air
temperature and pressure, respectively, for all stations. Dashed lines indicate
the average measured pressues; the top dotted black line is mean air
pressure at 2616 m for a standard atmospheric model (at 0 Celsius). Here day refers to all
samples taken while the sun was above the horizon, and night while below. The
strong near-sinusoidal variation of temperature and pressure with sun elevation
is evident, and the onset of fall and winter conditions is strongly correlated
with sun elevation falling below the horizon. A useful division between seasons
then is to refer to all samples taken when the sun is above the horizon as
``day" and all while below the horizon ``night."  In climatology studies this
would also be close to a clean division between fall-winter (sun down) and
spring-summer (sun up).

As the sun goes down, the winter low-altitude atmospheric thermal inversion
quickly develops over much of the High Arctic. It is a remarkably stable
phenomenon, intact throughout winter darkness. For Eureka, based on 50 years of
balloon soundings, the base of the inversion is almost always at the surface
from September through March, with a monthly median depth of about 800
m~\citep{Kahl1992}. At the peak of the inversion profile the temperature is
typically 8 to 14 degrees warmer than the surface, which effectively excludes
air below the inversion from rising and mixing with the free air above.
Interestingly, the lower quartile height during the winter at Eureka is always
below 600 m, that is, between 25\% and 50\% of the time it lies below the
elevation of PEARL. The upper monthly quartile in winter for Eureka is below
1200 m. The situation is almost identical at Alert, with inversion depths
shifted perhaps 100 m lower. Similar results are obtained for comparable Russian
locations~\citep{Serreze1992}, and MODIS satellite results~\citep{Liu2003}.

\begin{figure}
\plotone{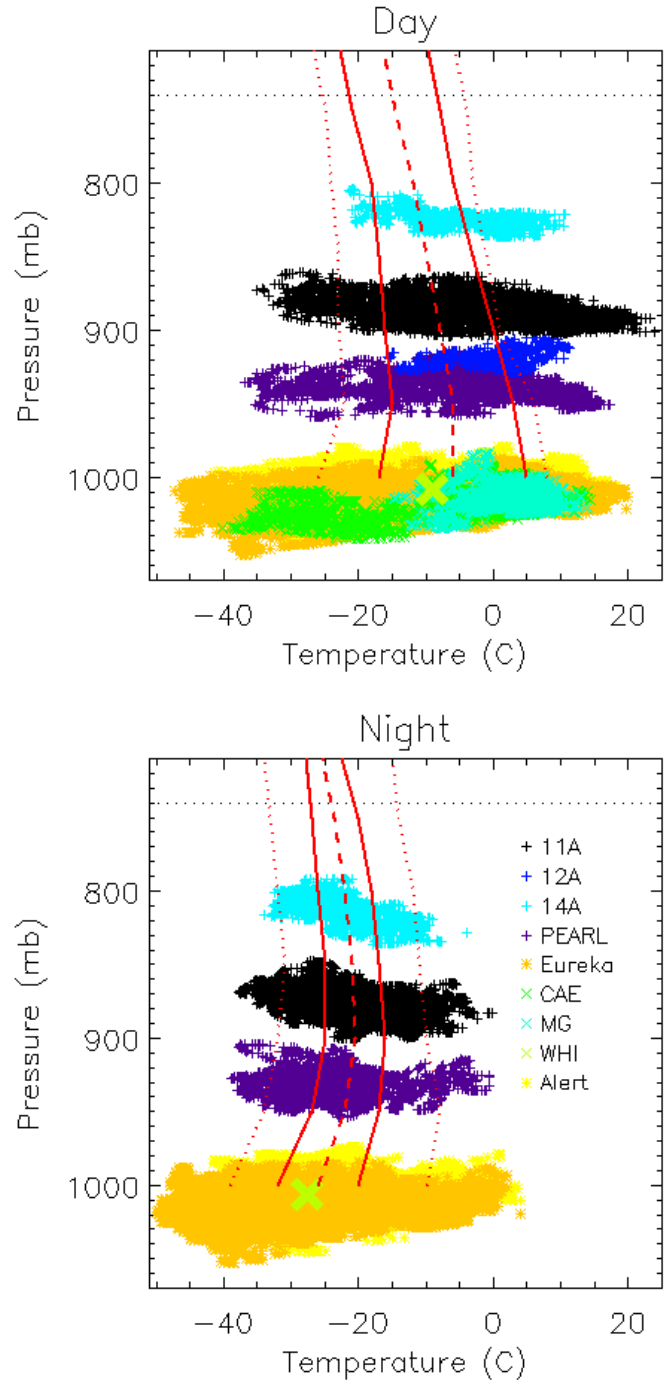}
\caption{Plot of all station barometric pressure data versus temperature, during
day (top) and night (bottom); day refers to the sun at or above
the horizon, and night to below. Solid red curves are
the average sping and summer (day) and winter and fall (night) temperature profiles at Eureka; dashed curves are the averages of day and night; dotted curves, quartiles.}
\label{figure_pressure_vs_temperature}
\end{figure}

The sharp contrast in air temperature profile relative to day and night is
evident in Figure~\ref{figure_pressure_vs_temperature}, where all station
temperature data are plotted relative to barometric pressure. Because barometer
data are missing for Ward Hunt Island, the mean temperature for day and night
are given at the mean barometric pressure for the same time periods at Alert,
which is the nearest with overlapping temporal coverage. Average sping/summer and winter/fall temperature profiles are shown for twice-daily Eureka aerosondes \citep{Lesins2009}. Assuming a consistent temperature profile along the coast, the data are consistent with Site 11A being above the inversion 75\% of the time or more at night, and Site 14A always being above the inversion peak.

\subsection{Prevailing Winds}\label{winds}

Wind speed and direction are given in the fourth and fifth from top panels in
Figure~\ref{figure_weather}. For the most part winds at all stations are calm.
This is punctuated by interruptions of one to a few days of high winds, with a
period of one or two weeks.

\begin{figure}
\plotone{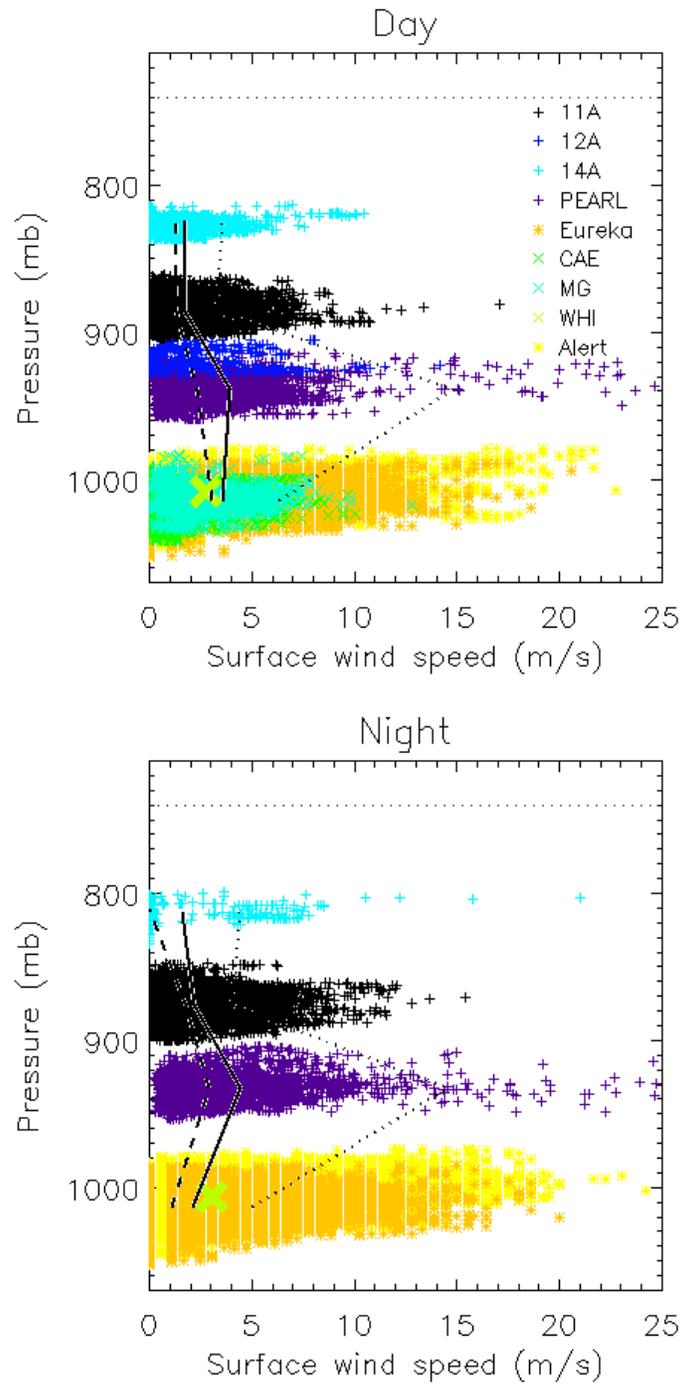}
\caption{Same as Figure~\ref{figure_pressure_vs_temperature}, except for surface wind speed. Dashed black lines connect the median pressures and wind speeds for Eureka, PEARL, Site 11A, and 14A; solid lines are means; dotted lines are those plus one standard deviation.}
\label{figure_pressure_vs_windspeed}
\end{figure}

\begin{figure}
\plotone{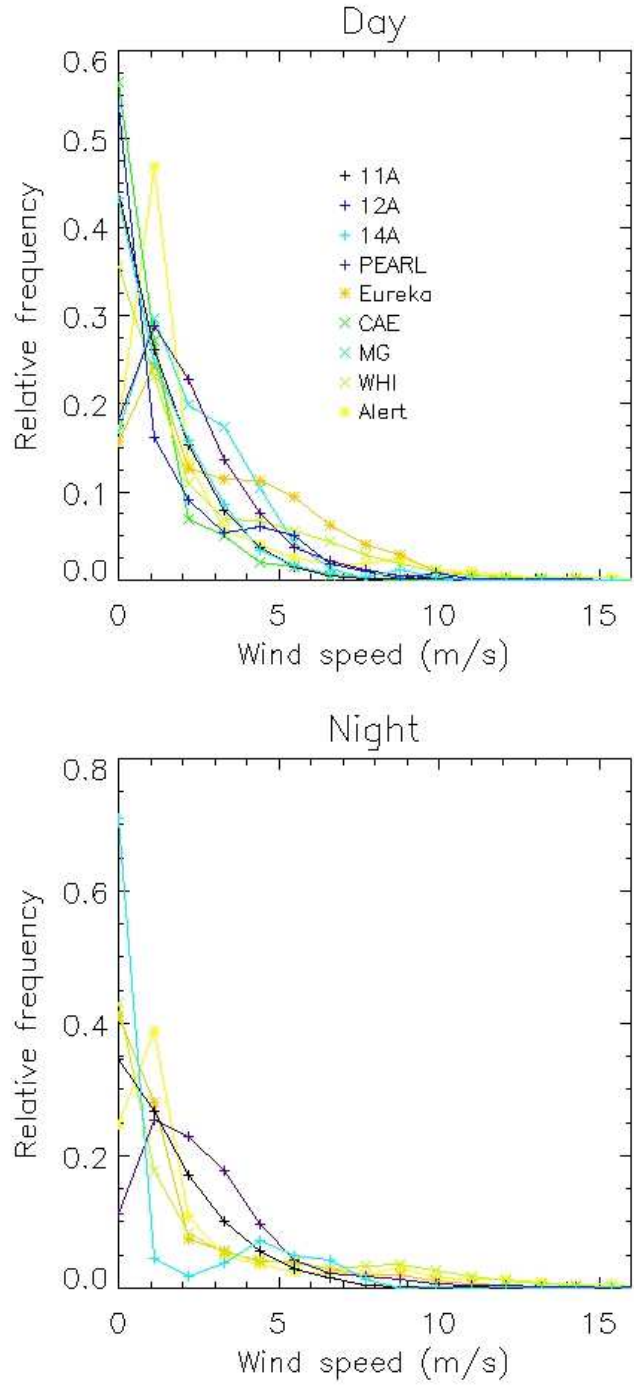}
\caption{Histogram of wind speed for all stations.}
\label{figure_histogram_windspeed}
\end{figure}

The difference between day and night wind speed is not as strong as for
temperature. Figure~\ref{figure_pressure_vs_windspeed} shows the wind speed for
all the stations relative to barometric pressure, separated into day and night.
One striking feature is the (relatively rare) occurance of very high
winds - over $20~{\rm m}~{\rm s}^{-1}$ at Alert, Eureka and PEARL. These do not
seem to be evident (at least in summer) at Cape Alfred Ernest or Milne Glacier,
nor as severe at the other coastal sites. The mean wind speeds for Eureka,
PEARL, Site 12A and Site 14A are shown as a connected line; dashed line, median;
dotted line, that plus one standard deviation. Median wind speeds are low, below
$3~{\rm m}~{\rm s}^{-1}$ at all stations, becoming less than $2~{\rm m}~{\rm
s}^{-1}$ at the higher sites. Figure~\ref{figure_histogram_windspeed} is a
histogram of wind speed for all stations, binned by units of $1~{\rm m}~{\rm
s}^{-1}$. In winter, it is evident that Site 11A is less windy than PEARL, and
Site 14A seems to be even calmer, although there is less data from which to
judge. For comparison, the median nighttime wind speed on Mauna Kea measured during site-testing for the Thirty Meter Telescope (TMT; 2m height at Site 13N) was $3.7~{\rm m}~{\rm s}^{-1}$ \citep{Schoeck2009}.

\begin{figure}
\plotone{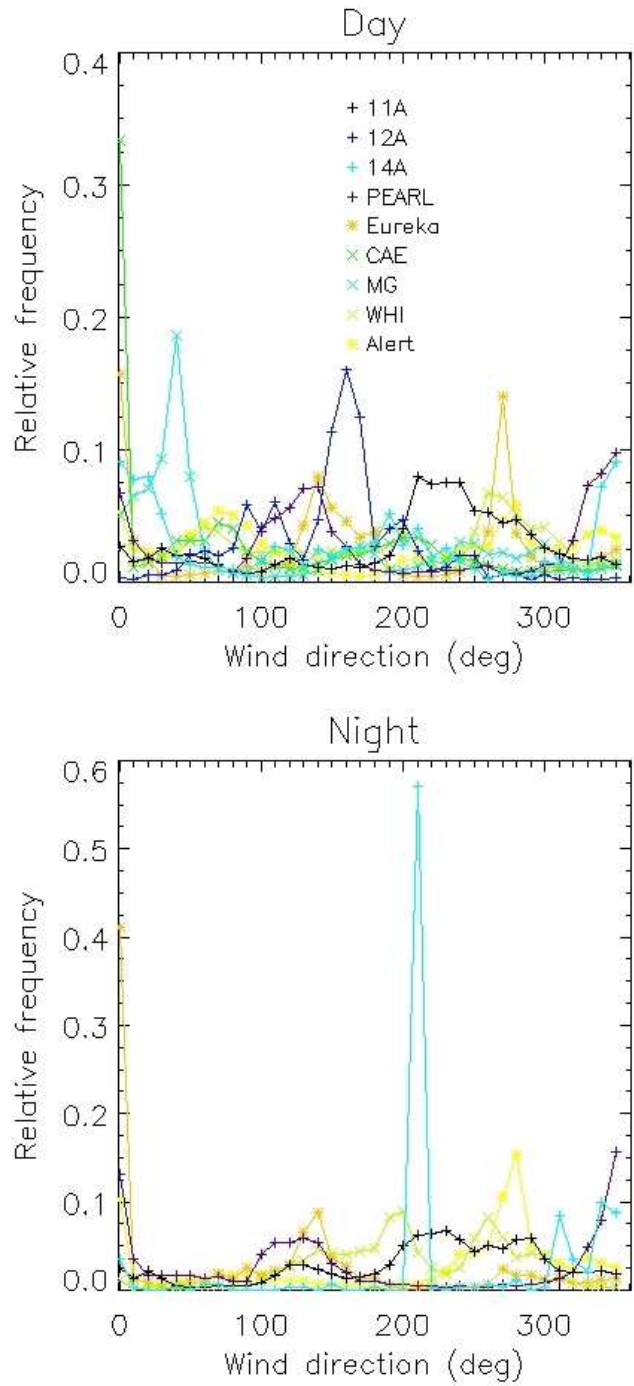}
\caption{Histogram of wind direction for all stations.}
\label{figure_histogram_wind_direction}
\end{figure}

Data from Alert indicates that the prevailing wind in the winter is almost always from the west, which
helped focus our attention on the western part of the island. Storms tend to come from the
southwest to west along the coast. Moreover, the Lake Hazen area is arctic
desert, in the precipitation shadow of the mountains to its
west. Eureka has a less obvious prevailing direction, perhaps even being
bimodal: south and west. Figure~\ref{figure_histogram_wind_direction} shows
histograms of wind direction binned in 10 degree increments, for day and night.
At Site 12A winds tend to be southwest to northwest in winter. The dramatic
spike in southwesterly winds at Site 14A is influenced by local terrain.
The saddle within which the station at Site 14A is located blocks winds from northwest and east to southeast.

\subsection{Clear-Sky Fraction}\label{clearsky}

\begin{figure}
\plotone{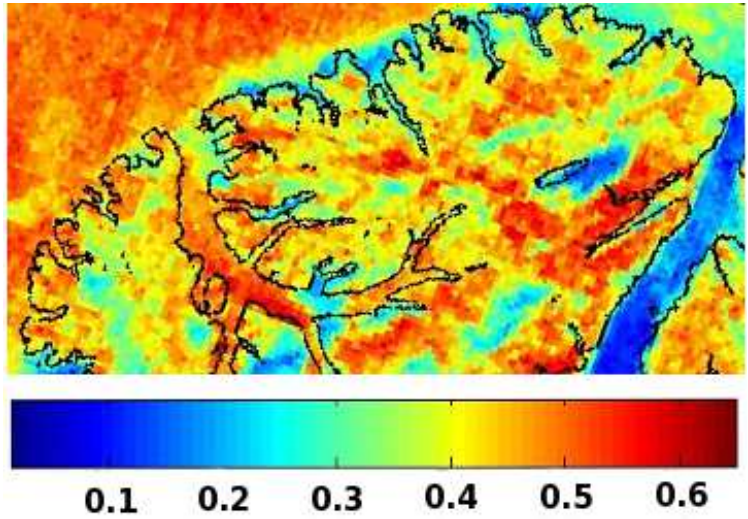}
\caption{Winter clear-sky fraction of time from MODIS data at 5 km resolution.
Flat low-lying regions, in the lee of mountains and between Ellesmere Island and
Greenland are especially cloudy. Best conditions, over 60\% clear-sky fraction,
correspond to the highest mountain peaks.}
\label{figure_ellesmere_sky_clarity_winter}
\end{figure}

To put our estimates of clear-sky fraction into context, a satellite analysis of sky
clarity over Ellesmere Island was first carried out. This was also useful in guiding our initial site selection process. Data from the Advanced Very High
Resolution Radiometer (AVHRR) sensor carried on the NOAA polar-orbiting
satellites is publicly available as atmospheric parameters via open-source
software (http://stratus.ssec.wisc.edu/products/appx/). This provides maps of
about 5 km resolution of winter clear-sky fractions, which are high for
northwestern Ellesmere Island. The results for 5 years of 11 $\mu$m data
beginning in 2000 are shown in Figure~\ref{figure_ellesmere_sky_clarity_winter}.

Characterizing sky clarity in nightime polar satellite images is challenging, because there is little contrast between thin cloud and snow. 
For Eureka, there has been some
success in showing from satellite images that ice crystals are blown from higher terrain
\citep{Lesins2009a, Bourdages2009}. And it is plausible that this effect is the
reason for the low clear-sky fraction (0.10 to 0.25) seen in Figure~\ref{figure_ellesmere_sky_clarity_winter} in the low-lying flat
terrain to the south of Eureka and near Lake Hazen, mid-way between Eureka and
Alert. Note also the poor visibility in the channel between Ellesmere Island and
Greenland, and right at the edge of the northern coast near our sites. At 5 km
resolution, this is blurred with pixels that include our sites (some within 5 km
of the coast), and so may give values of sky clarity that are too low. This is also
pessimistic for another reason, the algorithm used averages over several days,
and any thin clouds during that time are considered cloudy conditions during the
whole period. Higher spatial resolution data, coincident in time with ``ground-truth" observations, will be discussed later, but this preliminary satellite analysis
suggested a clear-sky fraction over 60\% for isolated mountains. Impressively, this may be better than Mauna Kea (at over 4000 m elevation), which based on SkyProbe data obtained on CFHT, has photometric conditions up to 60\% of the time \citep{Steinbring2009}.

The expectation of clearer skies at higher elevations, especially those above the peak of the thermal inversion, are confirmed with recent upward-looking Light Detection and Ranging (LIDAR) measurements from Eureka. These provide a vertical profile of clouds in winter - mostly ice crystals - from laser-backscatter cross-sections. Adopting an
optical depth threshold of 0.2 (what astronomers typically refer to as
photometric) the fraction of the time clouds were confined to below an altitude of 610 m (the elevation of PEARL) was under 10\%, growing to 13\% by 1000 m, and 17\% by 1500 m \citep{Lesins2009a}. The total integrated cloudy fraction was 33\% in these March 2006 data. Put another way, less than half the time that it was cloudy ($33/17$) were those ice crystals at altitudes greater than 1500 m.

\begin{figure}
\plotonenarrow{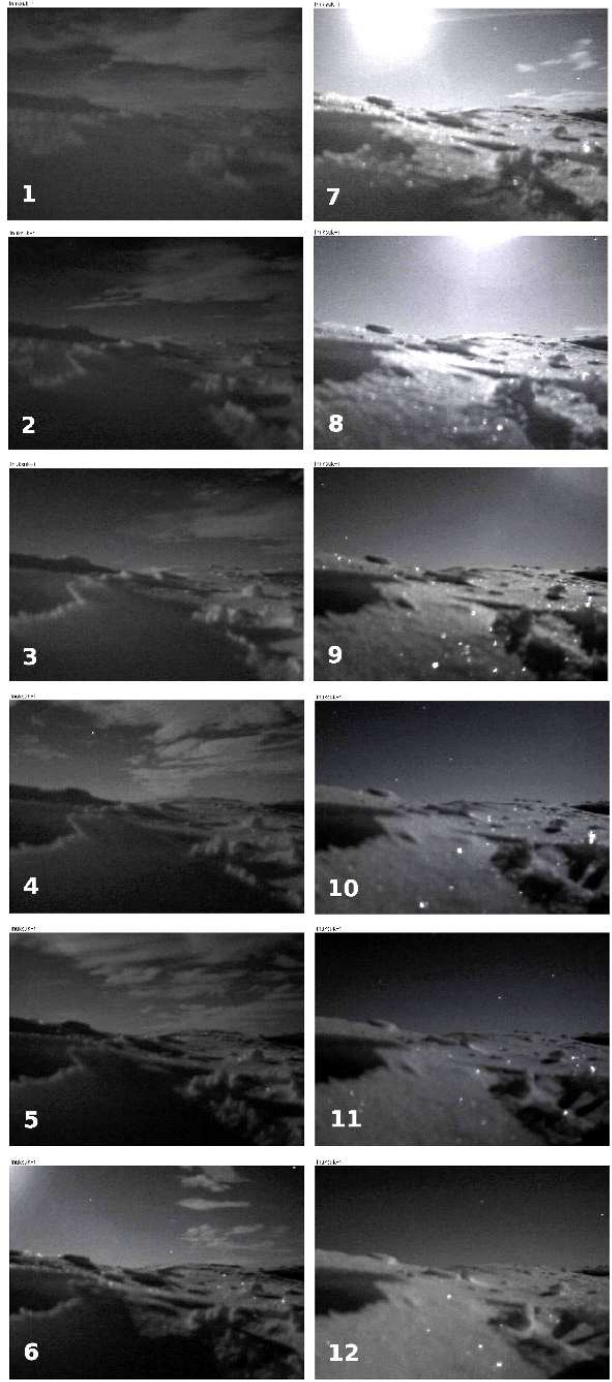}
\caption{Sample nighttime horizon-camera images from Site 11A; 12 sequential hourly images,
beginning with cloudy conditions (frame 1), becoming mostly cloudy (2 through
5), mostly clear (6 and 7), and finally clear (8 through 12). The moon is just
at the upper edge of the frame in frames 7 and 8; at least one bright star is
visible in frames 9 through 12.}
\label{figure_sample_horizon}
\end{figure}

All of our horizon camera images were analyzed following the procedures
discussed in \cite{Steinbring2008}. There were almost 2000 images taken at Site
11A, and a few hundred total at 12A and 14A. The wide field of view allowed good
discrimination of cloudy and clear-sky fractions during daylight hours. This
camera has an automatic iris, which opens fully under low illumination. Also,
the infrared cut-filter automatically retracts, providing imaging with the bare
detector. This provided good images until the onset of twilight, but for darker
conditions, only data taken when the moon was up and at illuminations greater
than 10\% were useful (see the top panel of Figure~\ref{figure_weather}). As the
moon is up for 24 hours at a time at these latitudes, much of those periods can be viewed in winter. 
Visual inspection was adequate for determining the conditions
of cloudy, mostly cloudy, mostly clear, and clear. See
Figure~\ref{figure_sample_horizon}, which shows a sample of 12 sequential hourly
images taken when the moon was up. Bright stars were also useful for determining sky 
clarity, allowing discrimination between a clear sky and uniformly flat overcast, especially if the stars can be seen progressing through the field in successive frames.
Some contextual information could also come from illumination of the foreground.
Note frames 10 through 12 in Figure~\ref{figure_sample_horizon}. The moon brightly reflecting off the snow is a strong indication that the sky is clear, in this case further buffered by stars being visible in the frames.

Visual estimates of cloud cover (in tenths of sky) are made hourly at Eureka.
The standard bins used by EC map almost exactly to our horizon-camera data: cloudy, mostly cloudy (over 5 tenths cloud
cover), mostly clear (less than 5 tenths, also designated by EC as mainly
clear), or clear. A condition of ``ice crystals" - often referred to as diamond dust - is not sensed by the horizon
camera, although any images of the moon with an obvious halo are deemed cloudy. It is common at Eureka under cold conditions and clear skies. All else
are ``indeterminant", which for horizon-camera data means either an ice-covered lens or electronically corrupted image, and for Eureka means any other condition besides cloudy, but likely to be so: rain, rain showers, drizzle, snow
pellets, snow, blowing snow, fog, freezing fog, snow grains, and snow showers. Eureka can be very cloudy, with conditions less than 30\% clear in
late summer. However, the intense cold and dryness of winter reduces cloud
cover. Water clouds become increasingly rare, with ice crystals dominating. The 50-year climatology for Eureka gives a mean cloudy fraction less than 50\% in November through March, so clear-sky fractions something like 50\% (some time will be partially cloudy) for most of winter \citep{Lesins2009}.

\begin{figure}
\plotonewide{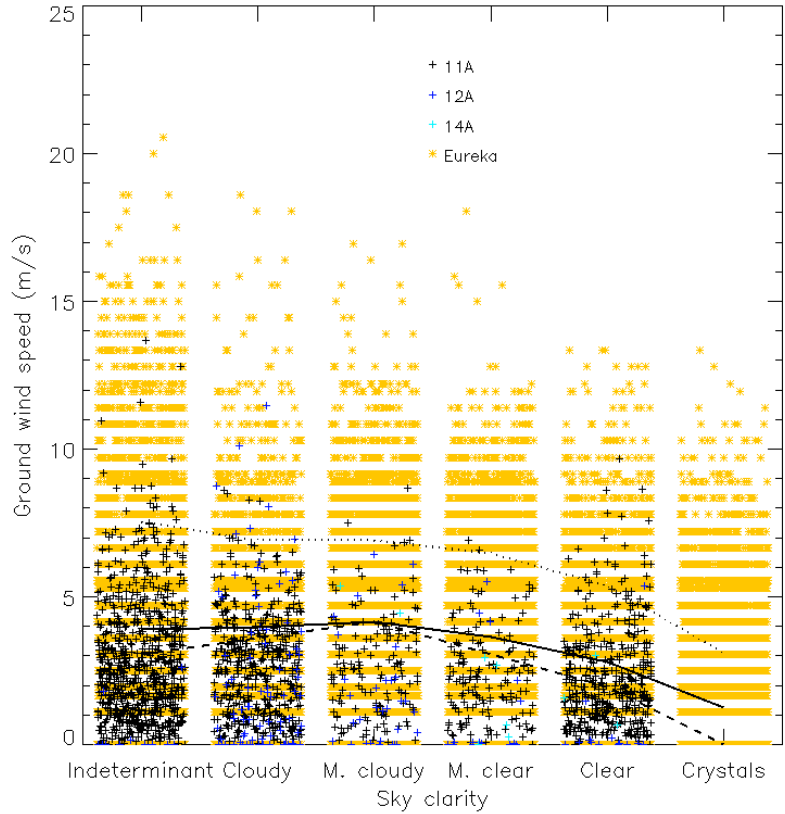}
\caption{A plot showing sky clarity measurements relative to wind speed for Eureka and
Sites 11A, 12A, and 14A; symbols are the same as in Figure~\ref{figure_histogram_wind_direction}. The condition of ``ice crystals" is not represented for the horizon-camera
data. Small random offsets have been applied to better show the distribution
wind speed. Overplotted are the median (dashed curve), mean (solid curve), and standard devation (dotted curve) for Eureka.}
\label{figure_windspeed_vs_clarity}
\end{figure}

\begin{figure}
\plotonewide{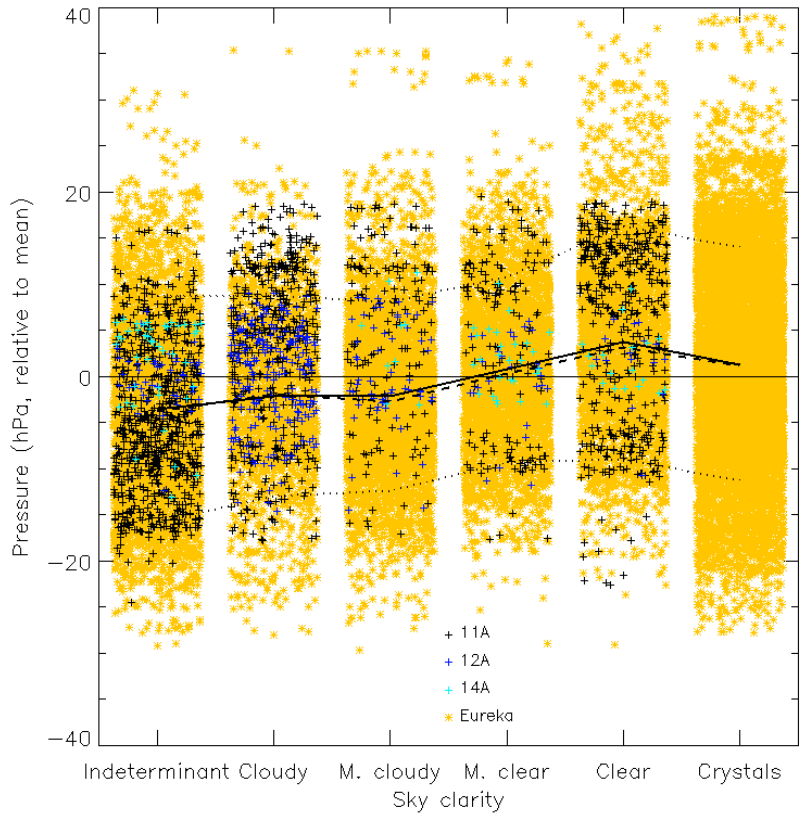}
\caption{Same as Figure~\ref{figure_windspeed_vs_clarity}, except relative to mean barometric pressure.}
\label{figure_pressure_vs_clarity}
\end{figure}

An issue for our wind-powered Inuksuit stations discussed in
\cite{Steinbring2008} is a bias towards obtaining horizon-camera observations during windy periods. That
this did not adversely affect our results can be demonstrated by
comparing with measurements from Eureka. Figure~\ref{figure_windspeed_vs_clarity} plots
all the horizon camera data against wind speed (average hourly wind speed nearest
in time to a sky observation). Small random shifts have been applied to help
show the distribution in wind speed. Also plotted are visual estimates at Eureka,
transformed to our bins. The median, mean, and standard deviation are overplotted. There is a strong correlation of low wind speeds and
clear skies, seen both in the Eureka data (which is not observationally biased to high wind speeds) and our horizon-camera data. Not surprisingly, clear skies are also correlated with high
barometric pressure. Figure~\ref{figure_pressure_vs_clarity} shows sky
clarity data plotted versus pressure relative to mean station pressure.

\begin{figure}
\plotonewide{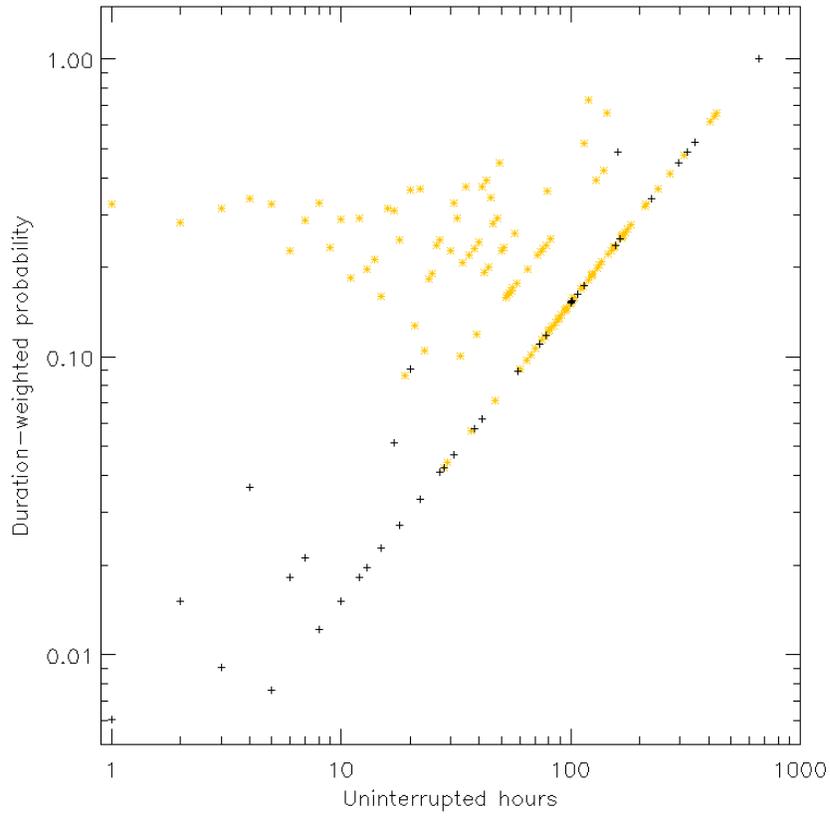}
\caption{Plot of duration-weighted probability of uninterrupted observations of
given number of hours during conditions considered mostly clear or clear for Eureka and
Site 11A. Eureka assessments include ``ice crystals",
which for higher elevations would likely be seen as clearer conditions.}
\label{figure_clear_duration}
\end{figure}

The duration of clear-sky periods is also comparable to, and perhaps better than Eureka.
For Site 11A, which has the most data of our sites, uninterrupted periods with either mostly clear or clear skies were tallied. The same was done for Eureka, with clear conditions including ice crystals. No observations are made at Eureka at midnight and 1 AM local time, and Site 11A has larger gaps. So all unobserved hours were assumed to have the conditions of the last observation. Each bin of the resulting histogram was multiplied by its associated duration. The results are shown in Figure~\ref{figure_clear_duration}, a plot of the
duration-weighted probability of an uninterrupted period of a given number of
hours. Site 11A and Eureka show similar trends, both having clear spans of a few hundred hours at a time, although the longest, at 659 hours on Site 11A (almost 4 weeks in May/June 2008) is possibly just a lucky string of shorter periods.

\begin{figure}
\plotone{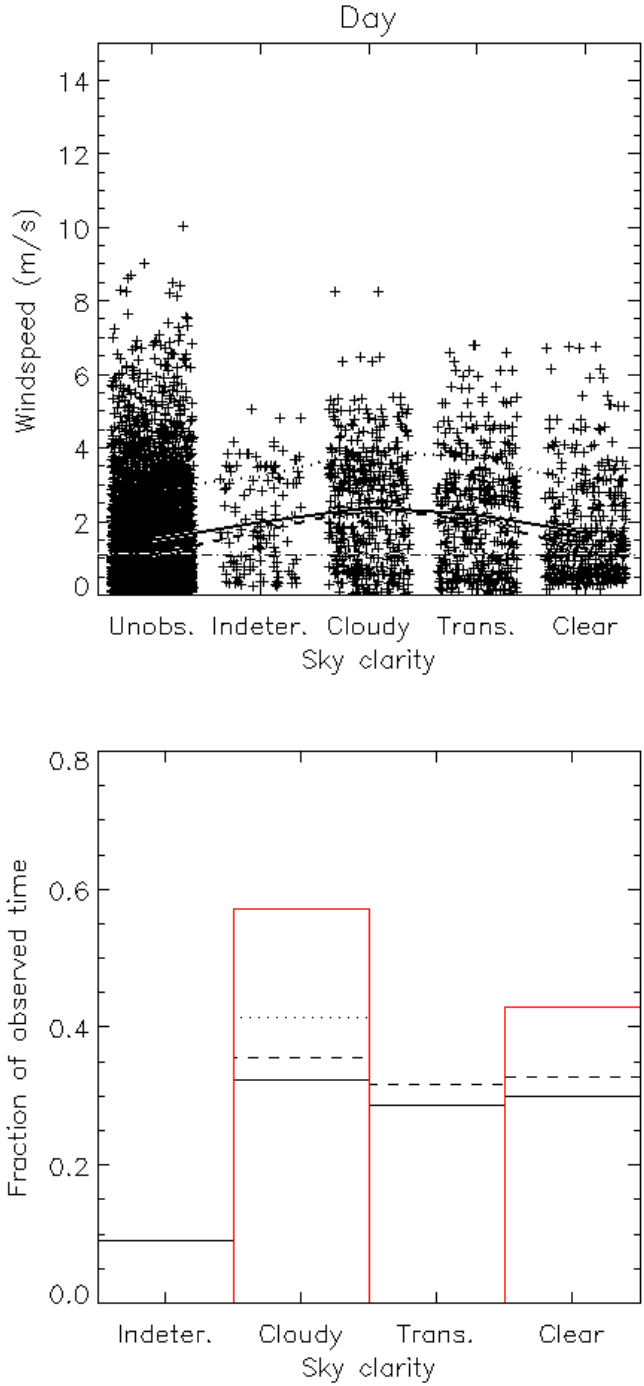}
\caption{Plots of sky clarity during the day for Site 11A relative to wind speed
(top) and fraction of observed time (bottom). Overplotted in the upper panel are
median wind speed (dashed line), mean wind speed (solid line), standard deviation (dotted line) for each bin, and global median (dot-dashed line). In the lower panel: redistributions of indeterminant samples for wind-speed bias (dashed line), assuming all indeterminants are cloudy (dotted line), and satellite results (red lines).}
\label{figure_sky_clarity_day}
\end{figure}

\begin{figure}
\plotone{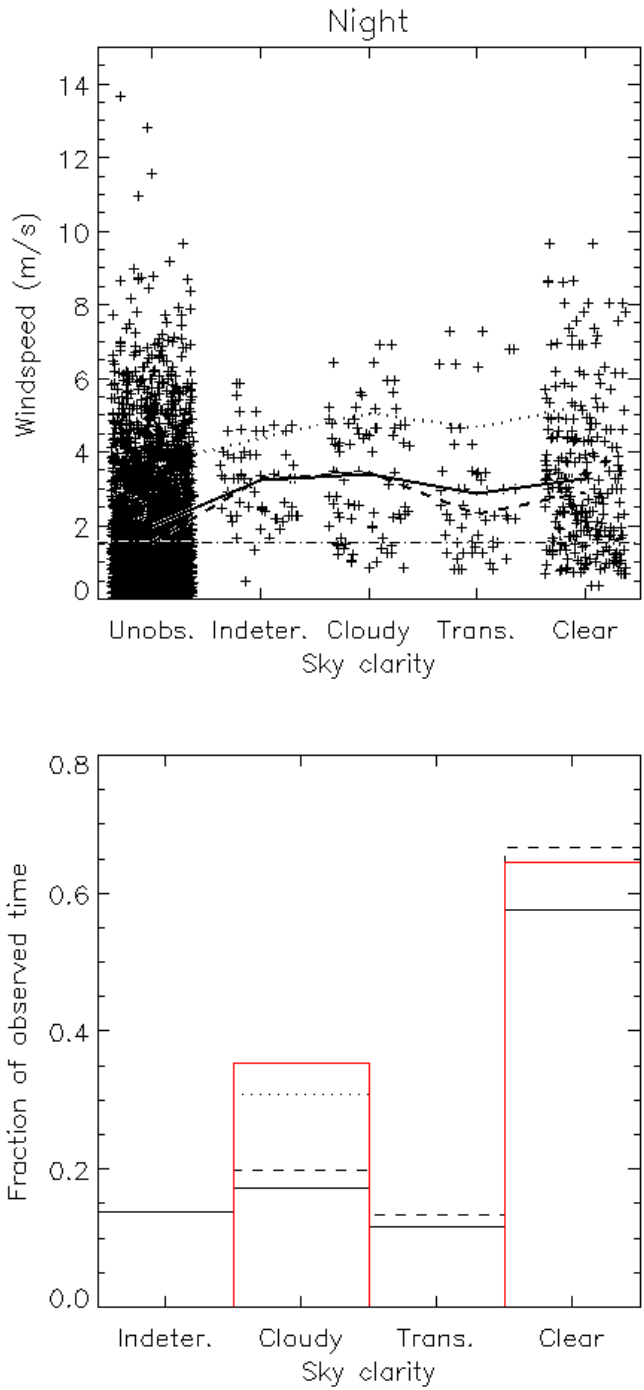}
\caption{Same as~\ref{figure_sky_clarity_day}, except at night. Data taken with moon below the horizon or illuminated less than 10\% are considered ``unobserved."}
\label{figure_sky_clarity_night}
\end{figure}

A direct comparison of horizon-camera clear-sky estimates at Site 11A and satellite imagery
was also carried out. The satellite analysis was repeated for better 1 km 
MODIS data with ground coverage of Site 11A when horizon-camera images were 
taken at similar times.
The resulting day and night clear-sky fraction from the Site 11A horizon camera are shown in Figures~\ref{figure_sky_clarity_day}
and~\ref{figure_sky_clarity_night}. Only data taken between fall 2007 and summer 2008 are used
here. There are many indeterminant measurements afterwards, and the timestamps are suspect (likely a problem with the
computer clock continually resetting) so potentially not accurate to within an
hour. The top panel shows the data relative to wind speed. Again, random shifts
have been applied to help show the distributions. The bottom panel shows the
fraction of observed time in each bin. For nighttime, only those data when the
moon was above the horizon and illuminated greater than 10\% are shown. To simplify the comparison with satellite analysis, which has just two bins: cloudy or clear, those horizon-camera observations designated mostly cloudy or mostly clear have been combined into one bin termed ``transitional."  Indeterminant observations (which may be cloudy, transitional, or clear) are accounted for in two separate ways. First, by redistributing them in proportion to median windspeed in the three other bins, which accounts for correlation of clearer skies with lower wind speed, shown as a dashed line.
Second, all indeterminant values were deemed cloudy, which is pessimistic, shown as a dotted line. For comparison,
the satellite result is shown in red. The horizon camera data seems to agree
with the satellite measurements; nighttime skies are clear 60\% to 65\% of the
time at Site 11A.

\subsection{Sky Transparency}\label{transparency}

Under clear skies, sky transparency should be very good on arctic mountain sites. A standard calculation
\citep{Hayes1975} suggests a mean atmospheric transparency at 1500 m due only
to Rayleigh scattering of 0.06 mag at $V$. An interesting possibility is a UV window due to ozone depletion \citep[e.g.][]{Brosch2009}. At lower elevations, one issue might be arctic haze, a phenomenon believed to be caused mostly by pollution
from Eurasia. It has a thin optical depth, in layers mostly restricted to below the inversion peak \citep[][]{Hoff1988, Quinn2007}, although this and ice crystals can persist at higher altitudes \citep{Lesins2009a}. For longer wavelengths, there is the benefit of little atmospheric water vapour. The cold air saturates, with relative humidities near 100\% throughout winter (second-to-bottom panel in Figure~\ref{figure_weather}), resulting in a low, stable water column. The 50-year monthly mean precipitable water vapor (PWV) from December through March
in Eureka is below 2 mm, only September and October are higher. This is already
comparable to the median measured by TMT on Mauna Kea: 1.9 mm
\citep{Otarola2010}. The February quartile in Eureka, when mean PWV reaches a
minimum, falls to 1.1 mm, more like the high Atacama plateau in northern Chile \citep{Giovanelli2001}. Eureka is at {\it sea level}, and a conservative estimate (scaling linearly with pressure) would suggest that Site
14A should have PWV below 1 mm most of the time in winter. Plans are already underway for
obtaining PWV measurements at PEARL using an IRMA device.

\subsection{Boundary-Layer Seeing}\label{seeing}

The ATP is designed for deployment at one of the mountain sites, but has
obtained some initial data during testing at PEARL. It is based on a lunar
scintillometer that UBC has operated successfully at Cerro Tololo (CTIO) for
several years, redesigned for operation in the harsh arctic environment. By
recording fluctuations in the lunar flux received by photodiodes over a range of
baselines, one can reconstruct the ￼$C_N^2$ profile, and from this determine the
seeing as a function of height above ground \citep{Beckers1993, Hickson2004,
Hickson2008, Tokovinin2010}. A similar device has also been operated on Mauna Kea (MK), and both it and the CTIO results checked against DIMM-MASS.

Close to 12 hours of data were obtained with the ATP, over about 70 hours of operation in fall and late winter 2009/10. Those results are discussed in
\cite{Hickson2010}, and briefly summarized here. The ground-layer seeing
statistics measured by the ATP at PEARL and the PTP at MK are very good. These
initial results indicate that the ground layer (GL) at PEARL is weaker
than at MK. The median GL seeing is 0\farcs28 compared to 0\farcs45 for MK. We
must also consider high-level turbulence not sensed by the scintillometer. For
MK, the median free atmosphere (FA) seeing is 0\farcs33 \citep{Schoeck2009}.
Adding this to the GL seeing values gives a median total seeing of 0\farcs46 for
PEARL and 0\farcs59 for MK. However, there is good reason to believe that the FA
seeing is better at PEARL than at MK, as MASS observations from Dome C in
Antarctica indicate unusually-weak FA turbulence \citep{Lawrence2004, Aristide2009}.
Like Dome C, PEARL is located inside the
polar vortex and may also have weak FA turbulence. In that case, the median
total seeing will be less than 0\farcs46, and lower still at heights greater than 6 m.

\section{Discussion}\label{discussion}

Our intial weather, sky-quality, and seeing data point to good conditions on
coastal Ellesmere Island mountains. A strong thermal inversion exists during
winter. Peaks over 1000 m are above the peak of this inversion
for much of the time. It is also dry, with high clear-sky fractions.
Compared to Eureka, clear-sky fractions are higher on coastal mountain sites, with the added benefit of calmer on-shore winds. This may also correspond with improved seeing, something we plan to test. There is already evidence from PEARL that the boundary layer at
these locations is weak. Impressively, for all basic site parameters: seeing, clear-sky fraction, PWV, and median wind speed, our results point to Ellesmere Island mountain sites being comparable to or better than Mauna Kea. Detailed characterization of transparency, sky brightness, and
seeing are planned for PEARL, although it can already been seen that there is a
compelling case for further investigation of the higher mountain sites. For
comparison, a summary of weather and site logistics is given in
Table~\ref{table_summary}.

\subsection{Site Practicalities}

At minimum, we have shown that we can access high-elevation coastal mountain
sites on Ellesmere Island. But as is typical in the arctic, changing
weather conditions can alter the best-laid plans very quickly. Not only fog, but
clouds at the summits have prevented us from reaching the sites. For example, 
helicopter landings have not been possible for roughly 70\% of the trips to Site 14A, leading to delay and increased expense for later return trips.
The snow-covered saddle presents no particular challenge; the peak itself is snow-covered too, but possibly workable. For the lower sites, which are on broad, long rocky ridges, helicopter access is more straightforward. Maneuverability around the sites themselves is easy. The ground is covered with boulders about 10 to 50 cm across, and slopes down some distance towards the sea at angles that can be climbed on
foot. Even so, a road to the base of either site is not practical, as both are
hemmed in by deeply crevassed glaciers. 

So far we have relied on a fieldwork plan mapped out with the PCSP field
managers that quickly gets us into and out of the field, with a short camp of a
few days. A Twin Otter carries all personnel and equipment to a predetermined 
unimproved airstrip near the sites. This is done when weather is known
to be good, both from satellite images and from reports from nearby camps. Camp
is then established, after which a Bell 206L (empty, apart from pilot) rendezvous
with us, coming from Eureka. About 3 drums of jet fuel are needed for a Bell
206L helicopter for 10 hours of flying time (including the trip back to Eureka).

As a base camp we have used two different airstrips over four field seasons. The
first is closer to the sites but also to the coast and
so more prone to fog. The pilots considered it somewhat short. The second 
is further away from the sites but less prone to fog. It is also
flatter, drier and longer than the other. Both have been made safer by filling
in some ruts with nearby rocks and marking out the edges with yellow wind
markers. A third has been overflown and is quite
wet in July, although it may be more workable early in the season, before the melt has progressed. Site 11A is
very close to this airstrip, and in fact, is within view from there.

Installing and maintaining equipment at the sites is no more difficult (or any
easier) than at other High Arctic locations. We have lost equipment to
storms, as have other groups. Communication via the Iridium satellite network has proved effective;
calls from the Inuksuit stations when powered (and the antennae operational) was
roughly once per week, sufficient to know health status. One route to improving
this to the level of obtaining data in real time might be to parallelize the
modems, as has been done at remote antarctic locations. Powering our small
stations and camera has been the most significant challenge, although we have
had reasonably good success using wind. A small fuel-cell electrical generator
was tried, with limited success. To get a sense of the feasibility of generating
sufficient power for a small telescope, a comparison can be made with the PLATeau Observatory (PLATO) site testing station at Dome A \citep[See][]{Hengst2008}. That uses a bank of
6 single-cylinder diesel engines burning jet fuel.
One is kept running at any time, for redundancy, and for waste heat to keep the
fuel from gelling in the cold. At its most efficient it can put out about 1 kW, and consumes
$300~{\rm g}~{\rm kW}^{-1}~{\rm hr}^{-1}$. For six months operation (assuming start-up at sunset) a similar system for us would require 
$300~{\rm g}~{\rm kW}^{-1}~{\rm hr}^{-1}\times 1~{\rm kW} \times 4020~{\rm hr} = 1206~{\rm kg}$, about what a Twin Otter can carry, and at least four trips up
the mountain with a Bell 206L. So delivering the fuel is comparable to what our
yearly logistics have been so far, and thus seems a reasonable task. Obtaining a
land-use permit to operate a diesel generator has not yet been investigated,
although it does not violate environmental regulations.

\subsection{Possible Development and Future Plans}

There is a well-defined process for obtaining permits for scientific studies in
Nunavut. Many of these physical studies involve extended camps of many
individuals, and the taking of wildlife or rocks and fossils, so a key aim of
the permitting process is to regulate the size of camps and environmental impact
on the study sites. Another is to respect Inuit land claims and
archeological sites. We have a scientific-study permit to test the three
mountain sites we have selected and access them from the camp. It is issued by
the Nunavut Research Institute (NRI) after review by the Nunavut Impact Review
Board (NIRB). This is renewed each year, which involves providing a
non-technical summary of the field work, along with a translation into
Inuktitut. Our permit is for testing the mountain sites and installing a small
telescope, and indicates an end to testing in 2012. It states that, if a larger
facility is considered, a new permit will be required.

All research projects funded by PCSP and granted scientific permits by the
Nunavut government are subject to screening for an environmental assessment by
NIRB. The regulations are the same as federal Canadian regulations. Our 
project so far falls in a category exempt from
assessment. First of all, we are not operating within a national park, which has
more stringent requirements. Also, the current stations are temporary, have a
footprint less than 25 square metres, do not involve the storage of 4000 litres
or more of fuel with 30 metres of a body of water, nor do they involve a field
camp exceeding 200 person days. Some other regulations which we are far from
exceding, such heights of communications antennae, could trigger assessments as
well. Our request for a scientific project permit renewal, after review by NRI,
is relayed to the local hamlets and hunter-and-trapper
associations (HTAs) for comment. For a site study this is sufficient, as our
work is very similar to other research in the area, is not near harvesting
areas, and so is not likely to raise concern. The purpose of relaying the permit
is to open dialogue, which we welcome. And to help this along, we are encouraged
to provide copies of published research to NRI. If a plan for a telescope moves
beyond the current feasibility study, a more pro-active approach to consultation
is warranted.

The already developed site at PEARL offers an alternative, compromising the best site quality in favor of established infrastructure. A competing site might be Pingarssuit Mountain (77N, 853 m) nearby Thule on Greenland. Solar observations made there in the late 1960s indicated at times ``excellent" seeing atop an abandoned radar dome \citep{Janssens1970} although there is no evidence from the literature that this was followed up with nighttime astronomy. At either of these ``low" sites, there are significant advantages to accessibility by road, and the availability of power, communications, and people. Arctic manpower costs are high though, so a telescope at a manned site would still likely be robotic, as it would necessarily be at a remote site. Any arctic telescope must be engineered to stringent levels of robustness and reliability. So at least for small telescopes, the main disadvantage of the remote sites is the need for deployment and maintenance by helicopter, which although tractable, is expensive.

We have presented an initial observatory site study of northwestern
Ellesmere Island, the first concerning arctic mountains. If a premier arctic site could be developed, it would be a compelling location for a
wide range of specialized infrared/optical telescopes. And in keeping with the accessibility of the sites in our study, some initial productivity may be
possible by deploying a modest telescope. For planet detection this might be
quite small, perhaps only 0.1 m in diameter. But this could quickly
demonstrate the value of investing in larger facilities, and a plan for a 2-m class telescope with adaptive optics is already being considered for PEARL.

\acknowledgements

A diverse group has joined forces on this effort, bringing together scientists,
engineers, and experts in arctic logistics. In particular, we thank Tim Hardy,
Kris Caputa, Murray Fletcher, Brad Wallace, Dell Bayne, Bruce Cole, and students
Mubdi Rahman and Johnathan Klein, the HIA ATRG-V, its shop and purchasing
department, and those of the Prairie and Northern Region of Environment Canada.
Data from the Cape Alfred Ernest, Milne Glacier, and Ward Hunt Island stations were kindly provided by Derrek Mueller, Luke Copland, and Warwick Vincent. Alert and Eureka meteorological data were provided by Environment Canada through the National Climate Data and Information Archive. Helpful comments came from Rene Racine, Tony
Travouillon, John Storey, Michael Ashley, and Ron Verrall. Our thoughtful referee, George Wallerstein, pointed out his original observations of refraction on ``P-Mountain" near Thule in the 1950s, and subsequent solar studies by others. We thank James Drummond and Pierre Fogal, and the technicians of the PEARL lab, who carried out setup and help operate the ATP and
SODARs. Logistical support has come from Environment Canada, which operates the
weather station at Eureka, as well as Natural Resources Canada through the PCSP
arctic-fieldwork base in Resolute Bay, for whose logistical support and skilled
pilots we are particularly grateful. This research was funded by the National Research
Council and the Natural Sciences and Engineering Research Council.

\newpage

\begin{deluxetable}{lccrr}
\tablecaption{Ellesmere Island Study Locations and
Comparisons\tablenotemark{a}\label{table_sites}}
\tablewidth{0pt}
\tabletypesize{\small}
\tablehead{\colhead{} &\colhead{} &\colhead{} &\colhead{Peak} &\colhead{Station}\\
\colhead{} &\colhead{Lat.} &\colhead{Long.} &\colhead{elevation} &\colhead{elevation}\\
\colhead{Name} &\colhead{deg N} &\colhead{deg W} &\colhead{m a.s.l.} &\colhead{m a.s.l.}}
\startdata
Mountain Sites     &     &     &     &\\
11                 &82.1 &83.4 &1720 &1098 (11A)\\
12                 &82.3 &82.3 &1402 &~777 (12A)\\
13                 &81.9 &80.7 &1696 &n/a \phantom{(11A)}\\
14                 &82.3 &80.5 &1868 &1639 (14A)\\
Mountain comparisons &   &     &     &\\
PEARL              &80.0 &85.9 &610  &616 \phantom{(11A)}\\
Barbeau Peak       &82.1 &83.4 &2616 &n/a \phantom{(11A)}\\
Coastal sea-level comparisons & & &  &\\
Eureka             &80.0 &85.9 &     &10 \phantom{(11A)}\\
Cape Alfred Ernest &82.3 &85.5 &     &10 \phantom{(11A)}\\
Milne Glacier      &82.1 &83.4 &     &15 \phantom{(11A)}\\
Ward Hunt Island   &83.1 &74.2 &     &10 \phantom{(11A)}\\
Alert              &82.5 &62.3 &     &31 \phantom{(11A)}\\
\enddata
\tablenotetext{a}{Ordered by longitude, west to east}
\end{deluxetable}

\newpage

\begin{deluxetable}{lcccc}
\tablecaption{Comparison of Sites\tablenotemark{a}\label{table_summary}}
\tablewidth{0pt}
\tabletypesize{\tiny}
\tablehead{\colhead{Property} &\colhead{PEARL} &\colhead{12A} &\colhead{11A}
&\colhead{14A}}
\startdata
\sidehead{Physical attributes}
Distance    &          &           &                       &\\
from coast (km)  &10   &2          &9                      &52\\
            &          &           &                       &\\
Elevation (m)    &616  &777        &1098                   &1639\\
            &          &           &                       &\\
\sidehead{Weather data and observed conditions}
Met. station     &     &           &                       &\\
samples\tablenotemark{b}   &8584   &1200     &17837        &3856\\
            &          &           &                       &\\
Sky clarity &          &           &                       &\\
samples\tablenotemark{c}   &\nodata  &470    &1983         &101\\
            &          &           &                       &\\
Min. time   &          &           &                       &\\
above peak  &          &           &                       &\\
inversion\tablenotemark{d} (\%) &25  &\nodata &75          &100\\
            &          &           &                       &\\
Median wind &          &           &                       &\\
at night (${\rm m}~{\rm s}^{-1}$)    &2.6  &\nodata  &1.5  &1.1\\
\sidehead{Practical and logistical issues}
Method of   &          &           &                       &\\
Access      &road      &helicopter &helicopter             &helicopter\\
            &          &           &                       &\\
Distance to &          &           &                       &\\ 
airstrip\tablenotemark{e} (km)    &15   &51         &31\tablenotemark{f}    &70\\
            &          &           &                       &\\
Time site   &          &           &                       &\\
accessible  &          &           &                       &\\
in summer\tablenotemark{g} (\%) &100 &80 &70  		   &30\\
\enddata
\tablenotetext{a}{Ordered by elevation}
\tablenotetext{b}{Hourly averages when all instruments operational}
\tablenotetext{c}{Once-per-hour assessments}
\tablenotetext{d}{Quartile relative to nighttime temperature profile at Eureka}
\tablenotetext{e}{For the remote sites, the further-inland airstrip}
\tablenotetext{f}{Although not used in this project, the nearest airstrip 
is 7 km}
\tablenotetext{g}{Estimated fraction of time that site was accessible to crew during field work, e.g., a planned trip not having to turn around
due to weather}
\end{deluxetable}

\newpage

\newpage

\begin{figure}
\plotone{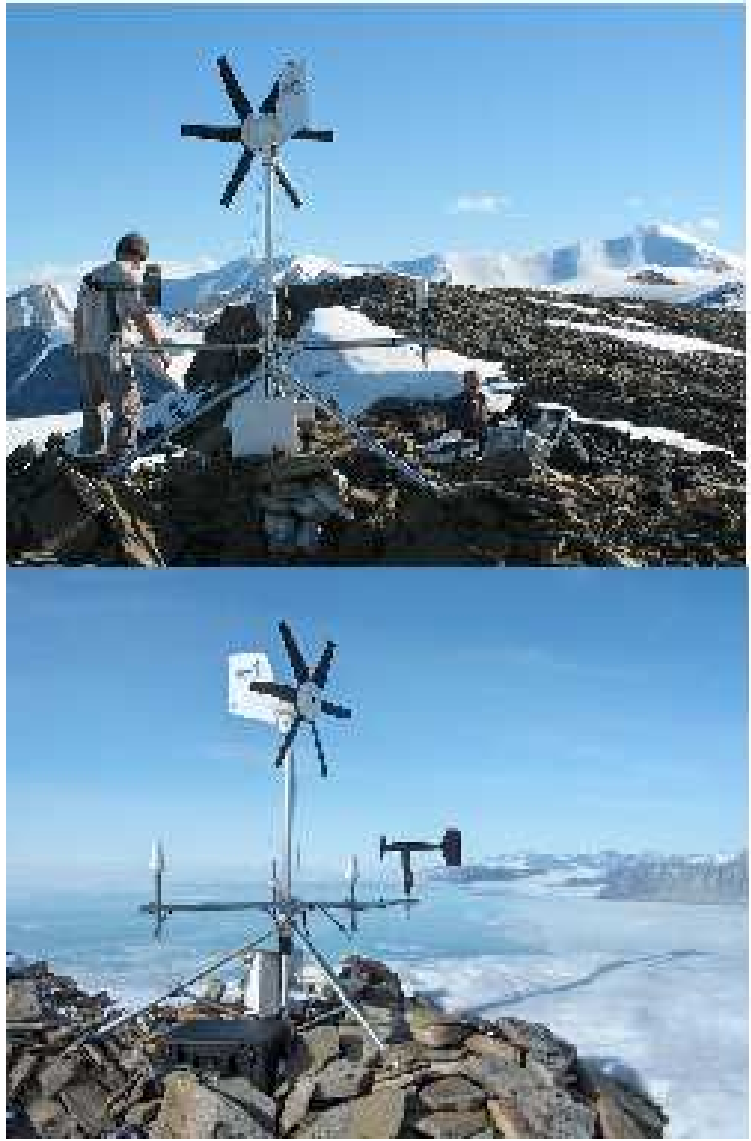}
\caption{Images of the Inuksuit station at Site 11A: facing south (top) and north (bottom). Interestingly, the two
photos are separated in time by three years; the top image was taken during the day of
deployment in 2006, the bottom during the latest maintenance visit in 2009. A thermometer housing is noticeably missing near the base of the wind-turbine tower.}
\label{figure_inuksuk1}
\end{figure}

\newpage

\newpage

\newpage

\newpage

\newpage

\newpage

\newpage

\newpage

\newpage

\newpage

\newpage

\newpage

\newpage

\end{document}